# Robust diffusion parametric mapping of motion-corrupted data with a three-dimensional convolutional neural network

Ting Gong[1], Qiqi Tong[1], Hongjian He[1], Zhiwei Li[2], and Jianhui Zhong[1,3]

[1]Center for Brain Imaging Science and Technology, Key Laboratory for Biomedical Engineering of Ministry of Education, College of Biomedical Engineering and Instrumental Science, Zhejiang University, Hangzhou, China

[2]Department of Instrument Science & Technology, Zhejiang University, Hangzhou, China

[3]Department of Imaging Sciences, University of Rochester, Rochester, NY, United States

## Abstract

Head motion is inevitable in the acquisition of diffusion-weighted images, especially for certain motion-prone subjects and for data gathering of advanced diffusion models with prolonged scan times. Deficient accuracy of motion correction cause deterioration in the quality of diffusion model reconstruction, thus affecting the derived measures. This results in either loss of data, or introducing bias in outcomes from data of different motion levels, or both. Hence minimizing motion effects and reutilizing motion-contaminated data becomes vital to quantitative studies. We have previously developed a 3-dimensional hierarchical convolution neural network (3D H-CNN) for robust diffusion kurtosis mapping from under-sampled data. In this study, we propose to extend this method to motion-contaminated data for robust recovery of diffusion model-derived measures with a process of motion assessment and corrupted volume rejection. We validate the proposed pipeline in two in-vivo datasets. Results from the first dataset of individual subjects show that all the diffusion tensor and kurtosis tensor-derived measures from the new pipeline are minimally sensitive to motion effects, and are comparable to the motion-free reference with as few as eight volumes retained from the motion-contaminated data. Results from the second dataset of a group of children with attention deficit hyperactivity disorder demonstrate the ability of our approach in ameliorating spurious group differences due to head motion. This method shows great potential for exploiting some valuable but motion-corrupted DWI data which are likely to be discarded otherwise, and applying to data with different motion level thus improving their utilization and statistic power.

# 1 Introduction

Diffusion weighted magnetic resonance imaging (DW-MRI) encodes information of the direction and speed of water diffusion in the intensity values of the acquired diffusion weighted images (DWI), which reflects the amount of hindrance experienced by water molecules along the applied direction of diffusion gradient owing to barriers and obstacles imposed by micro-structures [Assaf and Cohen, 2013; Jones et al., 2013]. By acquiring an adequate number of DWIs at different directions or strengths, diffusion models can be built to derive diffusion measures reflecting local microstructural tissue properties [Alexander et al., 2017a; Chang et al., 2017]. Being an informative technique, DW-MRI has been an essential tool that many worldwide large-scale projects put efforts on [Van Essen et al., 2012; Fan et al., 2016].

Over the last decades, the diffusion tensor imaging (DTI) has been widely used in brain maturation, development and aging studies [Qiu et al., 2015; Yoshida et al., 2013], and also to characterize neural tissue abnormalities in a variety of neurological and psychiatric disorders, including stoke [Thomalla et al., 2004], brain tumors [Inoue et al., 2005], neurodegenerative diseases [Cochrane and Ebmeier, 2013; Inglese and Bester, 2010; Stebbins and Murphy, 2009], schizophrenia [Kubicki et al., 2007], and autism [Travers et al., 2012]. However, the simplified assumption of Gaussian distribution in DTI model limits its ability to characterize more realistic heterogeneous tissue properties. Recent advanced models, including the diffusion signal model diffusion kurtosis imaging (DKI) and a variety of microstructural models [Assaf et al., 2008; Assaf and Basser, 2005; Zhang et al., 2012], have been developed to extend and complement the DTI-based techniques in numerous diseases [Van Cauter et al., 2012; Steven et al., 2014; Weber et al., 2015]. These advanced models, however, consist of increased model complexity, and therefore require substantially larger number of DWIs and applications of much stronger diffusion gradients, with consequently longer acquisition times and more stringent demand on hardware.

The demanding acquisition protocol for DW-MRI could cause practical difficulties. For instance, subjects are requested to stay still in MR scanner, but motion is inevitable [Elhabian et al., 2014a; Elhabian et al., 2014b]. When the number of DW scan is doubled, tripled, or even more, the data will be more susceptible to motion corruption, especially in scans of severely ill patients and young children, while usually those data are even more precious owing to consensus more difficult to obtain with those subjects. The corruptions of motion are of twofold: firstly slow motion causes misalignment between the DWI volumes, and secondly fast motion within a DWI volume results in signal phase shift or signal dropouts, altering the image intensity which cannot be fully recovered. These motion-induced effects can confound the measurement of interests. Previous studies using diffusion tensor imaging (DTI) has demonstrated increased uncertainty of model fitting process in motion-corrupted diffusion datasets, reporting overestimated fractional anisotropy (FA) in low anisotropy gray matter regions [Bastin et al., 1998; Landman et al., 2008] and diminished FA in high anisotropy white matter regions [Aksoy et al., 2008; Le Bihan et al., 2006; Jones and Basser, 2004] . This calls attention to possible bias in current findings from motion-corrupted DWI data both in individuals and groups of different motion levels.

There have been continuing efforts to minimize the impact of motion artifacts in diffusion MRI. The mostly used method is based on volumetric realignment assuming slow motion. Many image registration tools have been developed to mitigate the influence of this volumetric motion [Andersson and Sotiropoulos, 2016; Oguz et al., 2014; Rohde et al., 2004]. However, as mentioned above, within-volume motion also alters the image intensity, which cannot be corrected by realignment, and residual motion effects can lead to errors in the estimation of DTI derived measures [Ling et al., 2012; Liu et al., 2015; Oguz et al., 2014; Yendiki et al., 2014a]. To minimize the impact of motion on subsequent analysis, method detecting outlier of signal dropouts and replacing outlier with prediction [Andersson et al., 2016] can be helpful and should be used. Nevertheless, even following quality assurance and outlier replacements, increased extent of head motion difference between groups can still significantly impact the strength of the diffusion derived structural connectivity [Baum et al., 2018]. For studies of a group of subjects , rejecting data from subjects of excessive motion and adding motion parameters as nuisance regressors for the remaining subjects may be a compromised solution in maintaining the group-level statistical efficacy [Yendiki et al., 2014a]. For individual subjects, removing the corrupted volumes is feasible if there is data redundancy. However, rejecting volumes could cause detrimental impact during model fitting [Chen et al., 2015; Ling et al., 2012], and even introduce bias into the compared groups by different rejected number of volumes. Despite the investigations above, to the best of our knowledge, there still lacks of an effective tool to properly reutilize the motion-contaminated data and minimize the motion artifacts at individual level in diffusion MRI. And this is the main motivation behind the present study.

Recently, the deep-learning based methods have showed its ability in inference with high efficacy [Golkov et al., 2016; Lin et al., 2019; Zhu et al., 2018] in the field of MRI. We have previously proposed a three-dimensional hierarchical convolutional neural network (3D H-CNN) to recover all the eight DTI and DKI derived measures simultaneously from very sparsely-sampled DWI data [Li et al., 2019]. With this observation, we propose to extend this 3D H-CNN method to motion corrupted data after a motion assessment and volume rejection process. In this study, we evaluate and validate the performance of the proposed pipeline in recovering diffusion measures from motion contaminated data. The experiments were conducted on two different in vivo datasets, including a dataset of normal subjects with and without motion, and the other one of children with attention deficit hyperactivity disorder (ADHD) from the Healthy Brain Network Database [Alexander et al., 2017b]. The results suggest our method outperforms the conventional model-based fitting process in both two experiments substantially.

## 2. Materials and Methods

### 2.1 Datasets and scanning protocols

Two datasets of different imaging parameters and acquisition schemes were evaluated. The first dataset contains data from normal subjects scanned twice when asked to lying still and performing deliberate head motion, allowing a quantitative comparison to a reference for each individual subject. The second dataset of a group of children with ADHD was employed as a demonstration of the method in the vital research scenario of group level analysis.

**Dataset 1**

Four healthy adult subjects (S1-S4) were scanned on a 3T MAGNETOM Prisma scanner (Siemens Healthcare, Erlangen, Germany) equipped with a 64-channel head-neck coil. Two subjects (S1, S2) were scanned twice sequentially, when lying still for a reference standard scan and when performing deliberate movements for testing dataset. The other two subjects (S3, S4) were scanned once lying still constituting a training dataset. The detailed diffusion scanning parameters were as follows: single-shot echo-planar imaging (EPI) sequence; TR/TE=7000/67ms; FOV=210×210mm$^2$; slice number=50; resolution=2.5×2.5×2.5mm$^3$; slice acceleration factor=none; phase acceleration factor=2; phase partial fourier=none; bandwidth=2126 Hz/pixel; diffusion weightings of b = 1000, 2000, and 3000 s/mm$^2$ were applied in 30 directions, respectively, with six b=0 volumes entered, resulting in a total of 96 DWIs. The diffusion weightings and directions were designed using an uniform coverage across multiple shells and an incremental scheme by a generalization of electrostatic repulsion [Caruyer et al., 2013]. The b=0 volume with an opposite phase-encoding direction was also acquired, leading to a total acquisition time of twelve minutes for each scan.

**Dataset 2**

Data from 18 children diagnosed with ADHD (5 females and 13 males; age: 10.45±2.94 yr) were employed from the Healthy Brain Network Database [Alexander et al., 2017b]. The data were collected on a Siemens 3T Prisma scanner (Siemens, Erlangen, Germany) with 32-channel head coil. The detailed diffusion scanning parameters were as follows: simultaneous multi-slice EPI sequence; TR/TE=4100/89ms; FOV=240×240mm$^2$; slice number=81; resolution=1.7×1.7×1.7mm$^3$; slice acceleration factor=3; phase acceleration factor=none ; phase partial fourier=6/8; bandwidth=1700 Hz/pixel; one b=0 s/mm images and diffusion weightings of b = 1500, and 3000 s/mm$^2$ applied in the same 64 directions in each shell sequentially, resulting in a total of 129 DWI volumes. One b=0 image pair of reversed phase-encoding direction was also acquired. The total acquisition time for the diffusion session is about ten minutes for each subject.

## 2.2 Processing Pipeline

### 2.2.1 Preprocessing

The preprocessing steps were implemented in the FMRIB Software Library (FSL, University of Oxford, UK). For each DW-MRI scan, the non-DW image pairs with a reversed phase-encoding direction were firstly used to estimate the field map with the 'TOPUP' tool [Andersson et al., 2003]. Then the estimated field map was fed into the 'EDDY' tool, allowing an integrated correction of volumetric motion and both eddy-current and B0 field inhomogeneity induced distortions for the acquired DWIs [Andersson and Sotiropoulos, 2016]. The data was only resampled once in this procedure. To evaluate whether the correction method for replacing the within-volume motion-related signal dropouts attenuated the effects of motion to subsequent diffusion analysis, we also apply a command of replacing outliers in the 'EDDY' tool [Andersson et al., 2016]. Image slices in each diffusion encoded volumes that were detected with signal dropouts were taken as outliers and replaced by the non-parametric predictions about the expected signals, which are derived by a Gaussian process using angularly neighboring measurements. After correction, all DWI volumes were realigned to the first non-DW image volume and the corresponding DW gradient vectors were reoriented accordingly [Leemans and Jones, 2009].

### 2.2.2 Motion Assessment

To characterize the total head motion for each DWI in the scan, we derived measures of relative translation and rotation (RT, RR) to the first reference volume and absolute translation and rotation (AT, AR) to the consecutive volume, as well as the ratio of signal dropouts (RSD) based on slice outliers resulting from within-volume motion. After 'EDDY' in the preprocessing, six rigid transform parameters indicating relative translation and rotation from the first reference volume were obtained. The absolute translation and rotation components between each pair of consecutive volumes were then calculated. And slices in each diffusion-weighted volume that were detected as outliers were summarized by the 'EDDY' tool. The five motion measures for the $i$-th DWI in a scan are defined as follows:

$$RT_i = \sqrt{x_i^2 + y_i^2 + z_i^2}$$

$$RR_i = |\theta_i| + |\phi_i| + |\psi_i|$$

$$AT_i = \sqrt{(x_i - x_{i-1})^2 + (y_i - y_{i-1})^2 + (z_i - z_{i-1})^2}$$

$$AR_i = |\theta_i - \theta_{i-1}| + |\phi_i - \phi_{i-1}| + |\psi_i - \psi_{i-1}|$$

$$RSD_i = \frac{NO_i}{N_{total}} \times 100\%$$

where $x_i$, $y_i$ and $z_i$ are the three rigid translation components and $\theta_i$, $\phi_i$, $\psi_i$ are the three rigid rotation components around the X, Y and Z axes for the $i$-th DWI. $NO_i$ is the number of slices detected with signal dropouts for the $i$-th DWI and $N_{total}$ is the total number of slices in a volume. Since the first data volume is used as starting measure, all the components ($x_1, y_1, z_1, \theta_1, \phi_1, \psi_1, NO_1$) and measures ($RT_1, RR_1, AT_1, AR_1, RSD_1$) are zeros for the first reference volume.

### 2.2.3 Postprocessing

For conventional model reconstruction, a constrained weighted linear least square method was used for DKI reconstruction implemented in the Diffusion Kurtosis Estimator (DKE, The Center for Biomedical Imaging, Medical University of South Carolina, USA), which ensured physically and/or biologically plausible tensor estimates, increasing the robustness against noise, motion, and imaging artifacts [Tabesh et al., 2011]. For the proposed method, A 3D H-CNN designed for efficient DKI reconstruction [Li et al., 2019] was deployed to output all model-derived measures directly from the input DWIs. This H-CNN method relies on training data with training labels of good quality. Diffusion data from two subjects with full available DWIs were employed as the training data for network input and the model reconstruction results with the data were defined as the training labels. For motion contaminated data, thresholds of the five motion measures were set for selecting the remaining DWIs for analysis. The H-CNN were then trained using corresponding DWIs and training labels in the training data. The network was then applied to test data.

## 2.3 Experiments

All the experiments were performed under Linux operating system on a desktop computer with 20 Intel® Xeon® E5-2660 v3 @ 2.60 GHz CPU and a NVIDIA® Tesla k20c graphics card. FSL 5.0.10 and DKE 2.6.0 were installed for data preprocessing and model reconstruction. All the neural networks were implemented in Python using the Keras framework [Chollet and others, 2015] with a TensorFlow [GoogleResearch, 2015] back-end.

### 2.3.1 Experiment 1: Quantitative comparison for individuals

Experiment 1 was conducted on Dataset 1. The data were firstly corrected following the preprocessing process with and without signal dropouts replacement. For each subject at their still condition, all the 96 DWIs were employed for model-fitting, constituting the training labels (from S3 and S4) and reference standards (from S1 and S2). For the two tests at motion condition, different motion thresholds were set for selecting the less motion-affected DWIs to test the robustness of model fit and the H-CNN. The corresponding DWIs and the training labels from S3 and S4 were used to train the network, and the trained network was then applied to the test data. After reconstruction, the whole-brain root mean squared errors (RMSE) for each measure, including fractional anisotropy (FA), radial kurtosis (RK), mean diffusivity (MD), mean kurtosis (MK), radial diffusivity (RD), radial kurtosis (RK), axial diffusivity (AD), axial kurtosis (AK), and kurtosis fractional anisotropy (KFA), were computed and quantitatively compared to the reference at still condition.

### 2.3.2 Experiment 2: Group-level demonstration

Experiment 2 was conducted on Dataset 2. All data were preprocessed with signal dropout replacements. To summarize head motion for each subject, the five motion measures as defined in 2.2.2 were respectively averaged across all DWI volumes, and a total motion index (TMI) was calculated to integrate all motion measures into one motion score [Yendiki et al., 2014b], allowing a head motion ranking in a group according to the central tendency and dispersion of all the measures. The TMI for the i-th subject is calculated as follows:

$$TMI_i = \sum_{j=1}^{5} \frac{x_{ij} - M_j}{Q_j - q_j}$$

where $j = 1,…,5$ indexes the five average motion measures across all DWIs ($\overline{RT}$, $\overline{RR}$, $\overline{AT}$, $\overline{AR}$, $\overline{RSD}$), $x_{ij}$ is the value of the $j$-th average motion measure for the $i$-th subject, and $M_j$, $Q_j$, and $q_j$ are, respectively, the median, upper quartile, and lower quartile of the $j$-th average motion measure over all subjects included in the group comparison.

The 18 subjects were then divided into control Group.A with small motion and large motion Group.B according to its TMI. For the purpose of comparison, the conventional model reconstruction was firstly used, with all the DWIs reserved for diffusion kurtosis estimation, based on the finding that removing DW volumes would cause detrimental impact when estimating the diffusion measures [Chen et al., 2015]. For the 3D H-CNN method, data from two subjects with small motion were used as training dataset and the training labels were from the model reconstruction. For each subject, only DWIs that fulfill restriction of motion measures were reserved for testing and corresponding DWIs in the training dataset were used for training. After all reconstruction, voxel-wise statistical analysis of the FA data was carried out using Tract-Based Spatial Statistics (TBSS) [Smith et al., 2006] with threshold-free cluster enhancement [Smith and Nichols, 2009]. False discovery rate was used to correct for multiple comparisons with $p = 0.05$ as threshold for significance.

## 3 Results

### 3.1 Experiment 1

The motion measures of each DWI from the two testing subjects were depicted in Fig.1. In the 'still' condition (Fig.1a and b), the value of RT and RR were smaller than 3, AT and AR were smaller than 1, and outlier ratio was smaller than 6% for both subjects. The mean motion measures across all DWIs of $\overline{RT}$, $\overline{RR}$, $\overline{AT}$, $\overline{AR}$, and $\overline{RSD}$ were 0.78mm, 1.44°, 0.34mm, 0.93° and 0.56% respectively for S1, and 0.4mm, 0.74°, 0.41mm, 0.63° and 0.35% accordingly for S2. In the motion condition (Fig.1c and d), however, the $\overline{RT}$, $\overline{RR}$, $\overline{AT}$, $\overline{AR}$, and $\overline{RSD}$ were 0.83mm, 2.39°, 0.63mm, 1.13°, and 6.08% (1.06, 1.66, 1.85, 1.22, and 10.86 times respectively than in the still condition) for S1, and 2.75mm, 6.38°, 1.55mm, 3.30° and 6.52% (6.88, 8.62, 3.78, 5.24, 18.63 times respectively than in still condition) for S2. The number of DWIs reserved after different motion thresholds were listed in Table 1. There were 12 volumes for S1 fulfilling a restriction of RT<1mm, AT<1mm, RR<1.5°, AR<1.5° and RSD<5% and 10 volumes for S2 fulfilling a restriction of RT<3mm, AT<1.5mm, RR<3°, AR<2° and RSD<5%. S2 had a much more severe motion level than S1.

For each DWI, the large relative motion measures can be accumulated from the previous DWIs. However, when a large absolute motion occurs, it usually implies a quick motion and is very likely to cause within-volume deterioration. Fig.2 demonstrated two representative volumes from the two test subjects with large absolute motion measures after correction without and with outlier replacement. Visible image quality deterioration existed for both subjects without replacing outlier (Fig.2a and b). And for outlier replacement results, it showed some merit when the movements were comparatively moderate in S1 (Fig.2c). But when most of the volumes contains excessive motion, the replacement was not reliable such as in S2 (Fig.2d).

The whole-brain RMSE of reconstructed FA and RK and respective parametric maps against typical numbers of DWIs are show in Fig.3, using the motion thresholds listed in Table 1. For S1 with less severe motion (Fig.3a), the RMSEs of model fitting with and without outlier replacement had a consistent trend at all number of DWIs tested, and the outlier replacement provided a slight RMSE improvement of 0.005 and 0.015 on average for FA and RK respectively. The RMSEs for model fit were rather steady from 96 DWIs to 70 DWIs. After that, decreasing the number of DWIs caused a direct adverse effect to model fit with the RMSEs increasing from 0.094 and 0.285 with 70 DWIs to 0.141 and 0.398 with 27 DWIs respectively for FA and RK with outlier replacement correction. The H-CNN method was however rather robust against severe motion effects and the largely decreased DWI number, offering a steady performance and considerable improvement at all tested number of DWIs. Even with only 8 volumes retained, the RMSE was still 0.082 and 0.259 for FA and RK, offering improvement of 0.012 and 0.026 accordingly to the smallest achieved RMSE by model fit. For S2 with even more severe motion (Fig.3b), the outlier replacement with prediction can no longer provide apparent improvement for model fit with the RMSE curves nearly overlapped. And retaining more volumes with relatively large motions did harm to both model fit and H-CNN. Nevertheless, stable improvements were provided by H-CNN method, with the RMSE of 0.091 and 0.269 for FA and RK with 10 DWIs, compared to the smallest RMSEs of 0.138 and 0.350 with 65 DWIs for model fitting. These differences of RMSEs can be visualized from parametric maps. As shown in Fig.3c for S1, compared to the reference, the FA and RK maps with 70 DWIs of the smallest RMSEs showed a lower contrast with a blurred boarder. However, for the H-CNN with 8 DWIs, the fine structure was retained. This was even more apparent for S2 (Fig.3d) when a loss of structure was also prominent for model fit with 65 DWIs of the smallest RMSE. And a good contrast was still retained for H-CNN with 10 DWIs. The RMSEs for the other 6 measures of AD, MD, RD, AK, MK and KFA are shown in Fig.4. The RMSE tendencies from the two subjects were very similar to respective FA and RK measure for the H-CNN method, achieving stable and prominent improvements over the model fit. For S1 (Fig.4a), model fitting with outlier replacement provided improvements for the more complex measures (AK, MK and KFA) but not the diffusivity-related measures (AD, MD and RD). And there were no prominent differences for all the measures between model fit with and without outlier replacement for S2 (Fig.4b) with more severe motion level.

## 3.2 Experiment 2

The motion assessment results of each subject ranking by TMI are depicted in Fig.5. The average motion measures across all DWIs for all subjects ranged from 0.17-0.69mm for absolute translation, 0.08-0.87 degree for absolute rotation and 0.6%-5.12% for outlier ratio. Outlier replacement was employed for all subjects based on previous findings. For each subject, only volumes that fulfilled strict restrictions of RT<2mm, AT<1mm, RR<2°, AR<1° and RSD<5% were retained for H-CNN reconstruction based on previous assessment of subject in still condition and the accuracy of outlier replacement algorithm. The 7 Subjects with TMI<0 were grouped into small motion control Group.A (average absolute translation/ rotation/ outlier ratio/ number of rejected volume: 0.25mm/0.18°/1.27%/27) and the other 11 subjects were grouped into large motion Group.B (average absolute translation/ rotation/ outlier ratio/ number of rejected volume: 0.53mm/0.61°/2.89%/88) after manual inspection. Model and H-CNN reconstruction results for typical subjects are shown in Fig.6. For the subject in Group.A (Fig.6a), the FA and RK maps from model fit and H-CNN did not show prominent difference by visual inspection. However, for subjects with larger motion (Fig.6b), the maps from model fitting were blurry at tissue boundaries, while the H-CNN method achieved comparable contrasts to Group.A after rejecting a large number of volumes. The TBSS results of several group combinations are shown in Fig.7. In comparison between control Group.A and Group.B with large motion (Fig.7a), the model-reconstructed FA from Group B showed 71% voxels with significantly lower FA than the control group. This is in line with previous study that motion tends to decrease FA. However, for H-CNN method, no significant differences were found. To further investigate the difference between model and H-CNN reconstruction, paired t-tests between the two methods were performed (Fig.7b). For the control group, there was no significant difference found. However, for the motion group, 60.58% of the voxels for the model had significantly lower FA than that with the H-CNN method. These results imply that while motion deteriorates the model fitting, the H-CNN suffers little from the motion effect after rejecting motion-corrupted volumes.

## 4 Discussion

In deriving diffusion measures from DW-MRI, head motion has been realized as a confounding factor causing bias to the estimation. Previous works have suggested that residual effects are still present following standard head motion correction methods, and the removal of grossly affected DW images does not correct for that bias and may further introduce other detrimental effects. When utilizing the conventional model fitting method, our results were in line with the previous works. However, by applying our 3D H-CNN method, a significant improvement was achieved with minimal residual motion effects. Results on individual subjects with reference show that the proposed method provides accurate estimation of all DKI and DTI derived measures even with only 8 DWIs (from 96 DWIs) kept. The group analysis demonstrated the statistic power of the proposed method in robust parametric mapping for groups of subjects with different motion levels. Overall, the results suggest that our method has great potential to make full use of some valuable data from less cooperative subjects.

Theoretically, motion correction method can be divided into prospective and retrospective methods. Prospective methods can compensate motion effects by real-time motion tracking during the acquisition, reducing motion effects from its source. However, the use of such methods is limited by some external conditions, for example the need for an optical tracking systems [Aksoy et al., 2011], dependency on scanner platform [Maclaren et al., 2010], and/or increased acquisition time owing to reacquisition or calibration steps [Benner et al., 2011]. And the tracking accuracy is affected by noise in the navigator data. Hence the wide choice is still based on prevention of motion and the retrospective methods as additional image postprocessing steps. The most popular retrospective approach like what we use in this study relies on volume-to-volume registration to align all volumes to a reference [Andersson and Sotiropoulos, 2016]. There are also some more complex slice-to-volume registration methods which can reduce effects of motion within the acquisition period which causes zig-zag edges in an interleaved acquisition [Andersson et al., 2017]. This is more challenging and usually take lengthy processing time, especially for DWIs of high b value, which lacks anatomical features to be registered accurately. The registration methods do not correct motions occur during the diffusion encoding gradient causing signal dropouts. The outlier detection and replacement method we use can mitigate the detrimental impact of signal dropouts on subsequent diffusion measures estimation [Andersson et al., 2016]. However, the predictions that replace the outliers are in fact linear combinations of all the non-outlier data. The dropout signal will never be fully recovered. Hence the accuracy of the prediction will be affected by the number of remaining measurements. As show in Fig.3, outlier replacement does not provide much improvement for S2 due to its high outlier level. As our H-CNN method completely abandoned volumes with considerable motion, the method could directly avoid the introduction of other adverse effects. What is more important, the proposed method is not limited by any motion type and does not interfere with any of the correction methods mentioned above. Since our method can be the last defense being applied after other correction method, it may be the tool for solving situations where the motion is too severe to be accurately corrected by other means.

Rejecting motion-affected data as outliers as what we do can also be a retrospective approach for conventional model fitting, but this however requires data redundancy. It has been shown that the minimum number of distinct gradient directions necessary for robust estimation of FA value is around 30 [Jones, 2004]. However, for higher order DKI measures like the RK, MK, the estimation quality is closely related to the number of DWIs and multiple b-values, and the angular distribution of estimation precision is inhomogeneous [Sprenger et al., 2016]. Hence different rejection number in different subjects could introduce group bias. On the other hand, the H-CNN method benefits from the learning of large-scale training data, hence possessing a strong inference ability that could reduce the needed number of DWIs with very steady estimation performances. In our previous study, testing on 30 subjects from the human connectome project [Van Essen et al., 2012], the H-CNN method from 8 DWIs achieved statistically comparable estimation performance to model fitting with 96 DWIs using a reference of model fitting with 288 DWIs. This lays the foundation of rejecting outlier volumes without decreasing the estimation performance.

Though the proposed method is a retrospective method, there are implications from the view of acquisition. The first dataset we acquired employed an incremental scheme for b arrangement, i.e. diffusion vectors of different b-values were interspersed temporally, guaranteeing the available remaining data with multiple b-values even in an early-terminated scan. This is especially important for some microstructural models where data of multiple b-values are necessary. In addition, if the diffusion vectors are not coincident in different b-values, any subset data would more likely to result in a denser coverage of the angular space. These are very important aspects for model fitting process when fewer data are available, and might also benefit the network learning.

There are several limitations and possible improvements in the proposed model. It is noted that at least a full high-quality DWI training dataset is needed preceding the processing of any motion-contaminated data with the same imaging parameters, when the proposed method is used. This however should not be a major burden since it is typical to have a common protocol setup for a certain number of subjects in most clinical and large cohort studies. As long as at least one full dataset is available, the network can be trained and applied to any possible motion-affected patterns (different combinations of affected DW directions and b-values) just like what we have done with dataset 2. Moreover, this limitation could possibly be resolved by diffusion data simulations [Graham et al., 2016], which is an ongoing work we are making effort with.

In our current study, we give quantitative comparison results for individual subject in experiment 1 with data acquired for the study, and group level statistical demonstration in experiment 2 with public data. Though the results showed prominent improvements of the proposed method in both experiments, the number of subjects and the investigated cases in experiment 2 for group demonstration are limited. It is not easy to find a publicly available database containing multi-shell diffusion data with varying data qualities from motion-prone subjects. In this regard, the Healthy Brain Network Dataset we employ is a very comprehensive platform as it provides datasets of varying quality without interfering and handling possible motion during acquisition from children with various types of illness. This could help to facilitate the development of artifact correction strategies and the evaluation of impact of real-world confounds on reliability. Hence, we exploited the data from 18 children of the same phenotype of ADHD, which is the largest category from the various phenotypes in their old data release from one of the 3T acquisition center. We plan to gather a larger dataset in our future study, allowing more detailed exploration of the method in group research scenario.

The other possible improvement for our method is to extend our work to more diffusion models. The current study has evaluated all the diffusion kurtosis and diffusion tensor derived measures, which are the most widely used diffusion metrics in diffusion MRI. However, there are other informative diffusion models benefitting from the combination of multi-shell protocols with high angular resolution, and could also face the challenge of head motion from the increased acquisition time. Examples include the neurite orientation dispersion and density imaging (NODDI) [Zhang et al., 2012] which provides metrics reflecting specific microstructural tissue properties, and model of fiber orientation distributions [Jeurissen et al., 2014] which can reflect brain structural connectivity. Future studies taking those models into considerations will further test the feasibility of our CNN-based pipeline for robust diffusion estimation.

## 5 Conclusion

In this study, the proposed 3D H-CNN method was used for recovering diffusion measures robustly from motion-contaminated diffusion data, and validated with two different datasets showing both its quantitative and statistical benefits. By motion assessment and rejection, the H-CNN method could be a new powerful tool in handling motion-contaminated data, providing increased utilization of valuable data from less cooperative subjects such as certain motion-prone patients, young children and infants.

## Acknowledgements

We acknowledge the supports from National Key R&D Program of China (2017YFC0909200), National Natural Science Foundation of China (81871428, 91632109), Shanghai Key Laboratory of Psychotic Disorders(13dz2260500), and Major Scientific Project of Zhejiang Lab (No. 2018DG0ZX01), the Fundamental Research Funds for the Central Universities(2019QNA5026).

# Tables

Table 1. The number of DWIs reserved (Ndwi) with different motion measure thresholds for S1(a) and S2(b).

(a) S1

| Ndwi | 96 | 90 | 82 | 70 | 57 | 52 | 49 | 39 | 27 | 16 | 12 | 8 |
|---|---|---|---|---|---|---|---|---|---|---|---|---|
| RT/mm | 2.5 | 2 | 2 | 2 | 2 | 1.5 | 1.5 | 1.5 | 1.5 | 1 | 1 | 1 |
| AT/mm | 2.5 | 2 | 2 | 1.5 | 1.5 | 1.5 | 1.5 | 1.5 | 1 | 1 | 1 | 0.5 |
| RR/° | 6 | 5.5 | 5 | 4.5 | 4.5 | 4 | 3.5 | 3 | 2.5 | 2 | 1.5 | 1.5 |
| AR/° | 5 | 4.5 | 4 | 3.5 | 3.5 | 3 | 2.5 | 2 | 2 | 1.5 | 1.5 | 1 |
| RSD | 25% | 20% | 15% | 12% | 10% | 10% | 10% | 10% | 5% | 5% | 5% | 5% |

(b) S2

| Ndwi | 96 | 91 | 85 | 79 | 69 | 65 | 49 | 40 | 24 | 14 | 10 | 6 |
|---|---|---|---|---|---|---|---|---|---|---|---|---|
| RT/mm | 6 | 5.5 | 5.5 | 5.5 | 5 | 5 | 4.5 | 4.5 | 4 | 4 | 3 | 2 |
| AT/mm | 4.5 | 4 | 4 | 4 | 3.5 | 3.5 | 3 | 3 | 2.5 | 2 | 1.5 | 1.5 |
| RR/° | 14 | 13 | 12 | 11 | 10 | 9 | 8 | 7 | 6 | 5 | 3 | 2 |
| AR/° | 11.5 | 11 | 10 | 9 | 8 | 7 | 6.5 | 6 | 5 | 4 | 2 | 1.5 |
| RSD | 60% | 50% | 40% | 30% | 20% | 15% | 10% | 10% | 8% | 8% | 5% | 5% |

# Figures

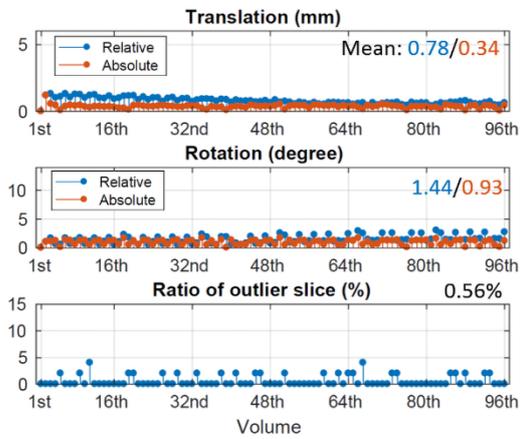
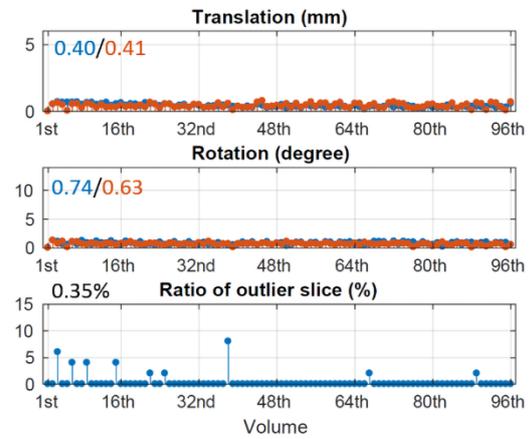
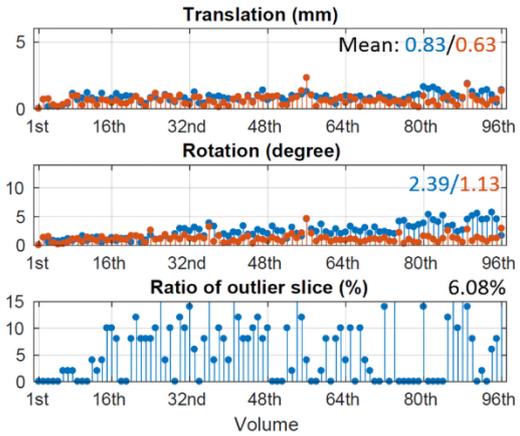
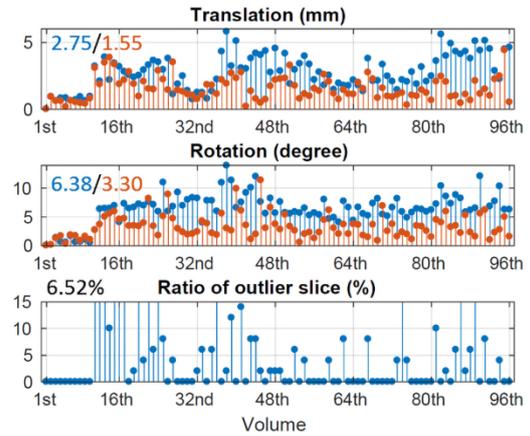

Fig.1 Demonstration of the five motion measures of the 96 DWIs from the two test subjects in still and motion condition respectively. The mean motion measures across all DWIs were computed and depicted in the figure.

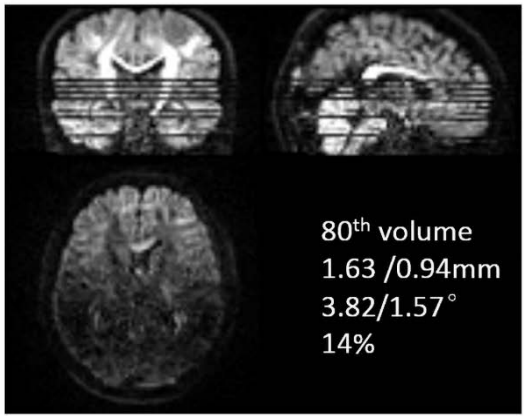 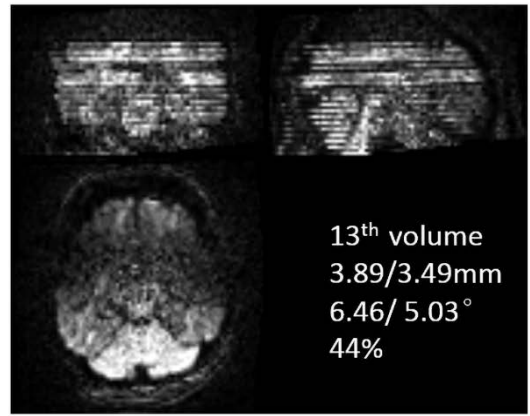
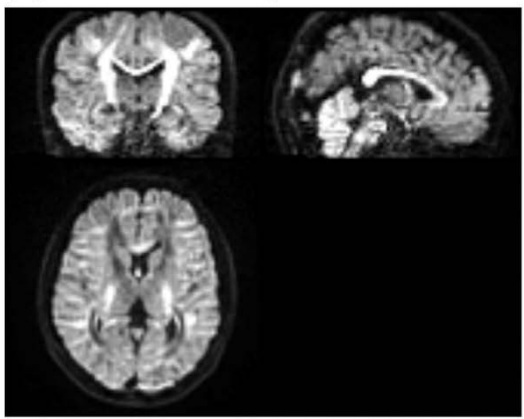 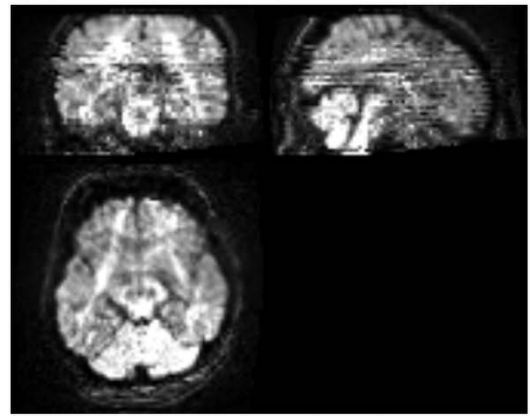

Fig.2 Typical volumes with large absolute motion measures after preprocessing without (top) and with (bottom) outlier replacement. (a)(c) the 80[th] volume from S1 and (b)(d) the 13[th] volume from S2. The motion measures were depicted in the figure with the order of RT/AT in mm, RR/AR in degree, and RSD in percentage.

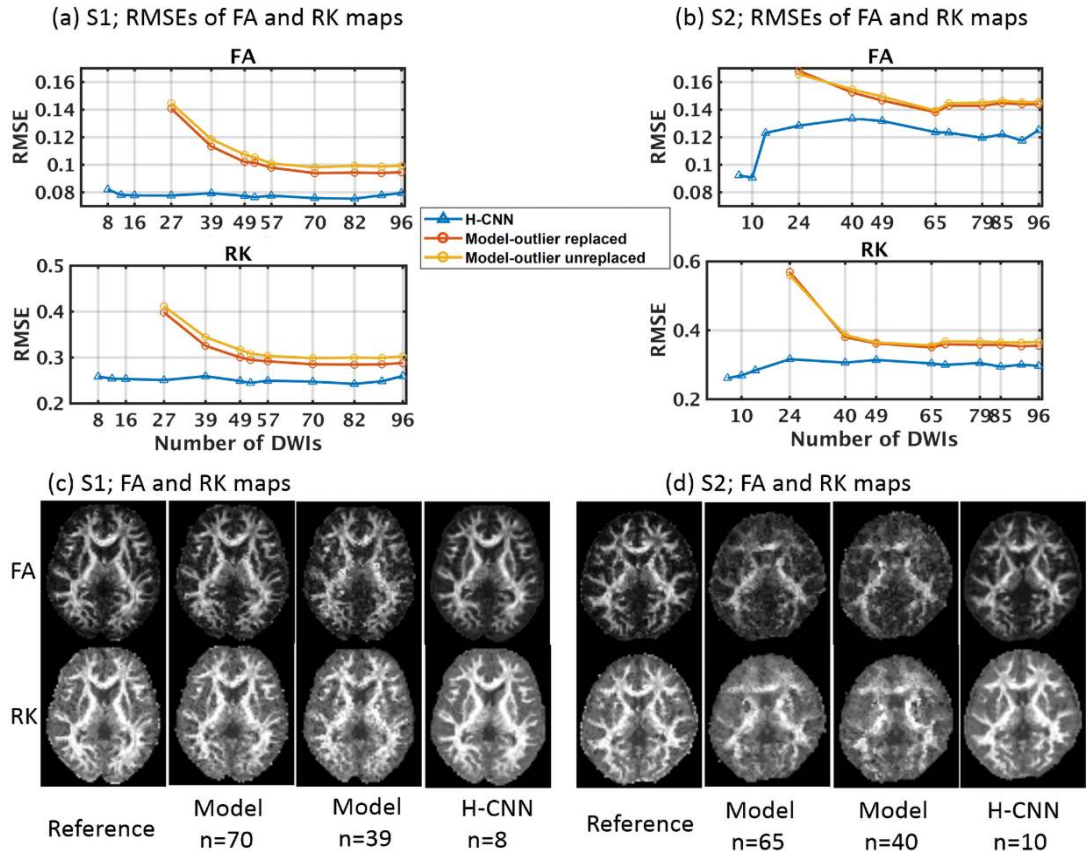

Fig.3 The whole-brain RMSE as a function of the number of DWIs selected by different motion thresholds (a)(b) and reconstructed maps for FA and RK (c)(d). The model-fitting results with full dataset (96 DWIs) in the still condition were used as a reference to compute RMSE.

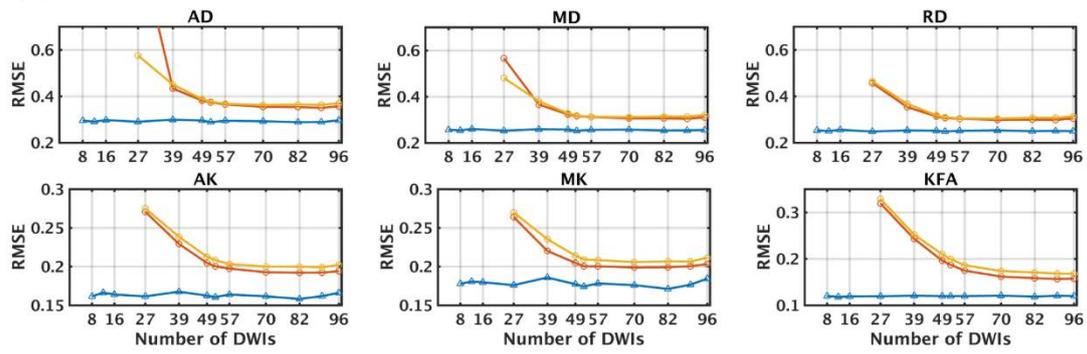
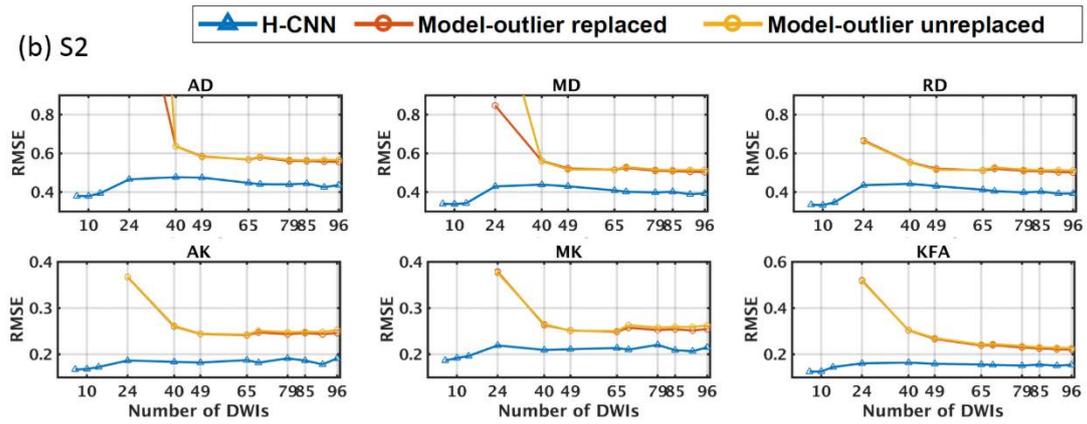

Fig.4 The whole-brain RMSEs for the other 6 measures from S1(a) and S2(b), including AD, MD, RD, AK, MK and KFA.

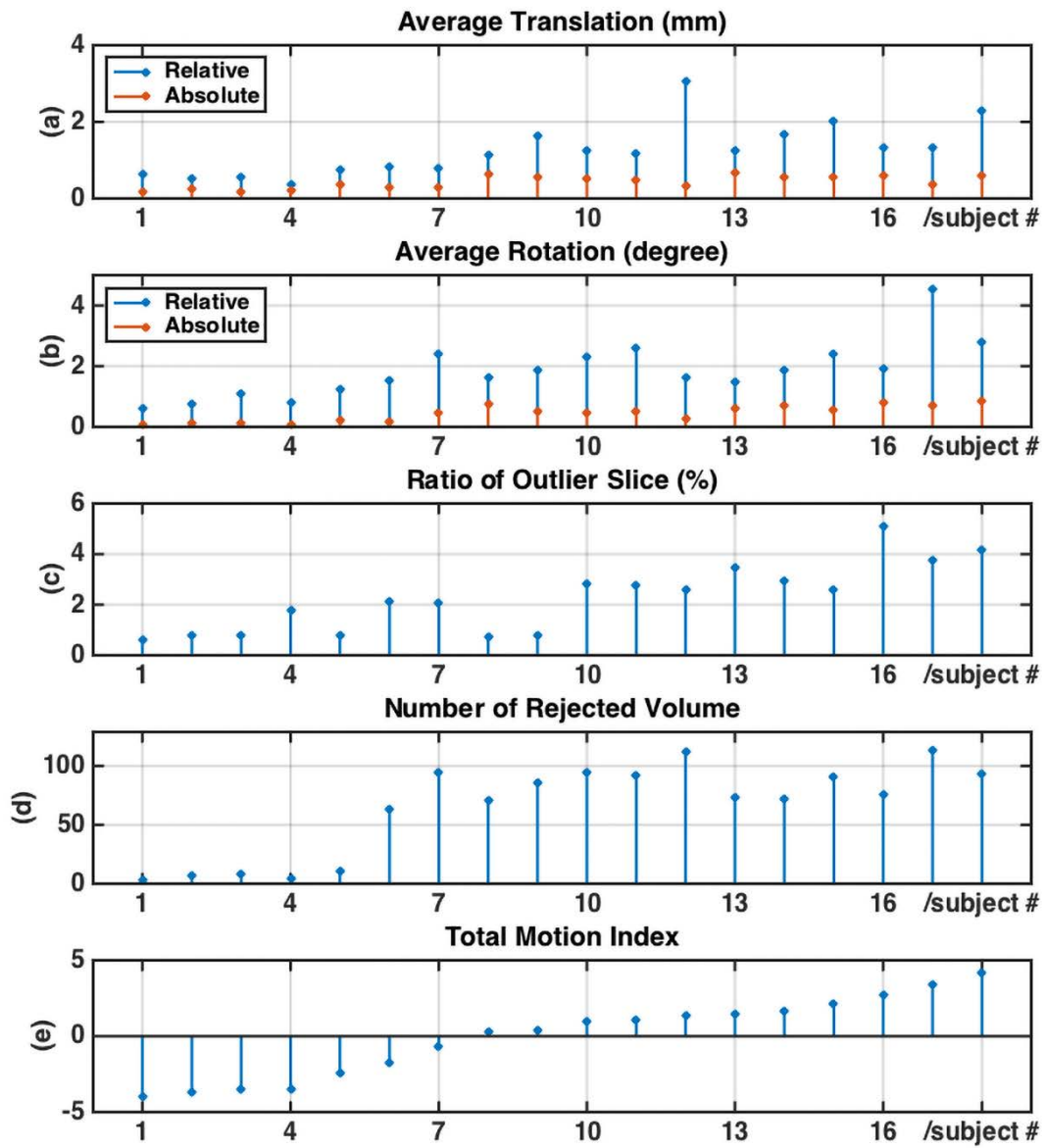

Fig.5 Motion assessment results for each subject. (a)-(c) The averaged 5 motion measures across all DWI volumes. (d) Number of rejected volumes for H-CNN reconstruction using a criterion of RT<2mm, AT<1mm, RR<2°, AR<1° and RSD<5%. (e) The TMI calculated taking the above 5 measures into considerations. The 7 Subjects with TMI<0 were grouped into small motion control Group.A and the other 11 subjects were grouped into large motion Group.B after manual inspection.

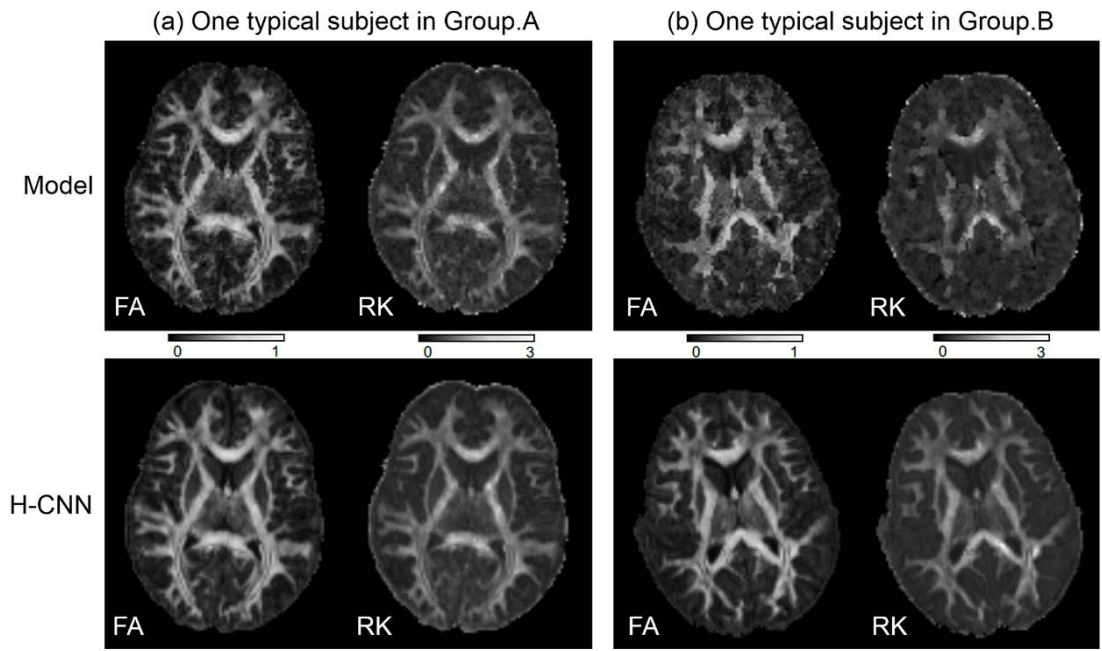

Fig.6 Model and H-CNN reconstructed FA and RK maps of a typical subject in Group.A (a) and another in Group.B (b).

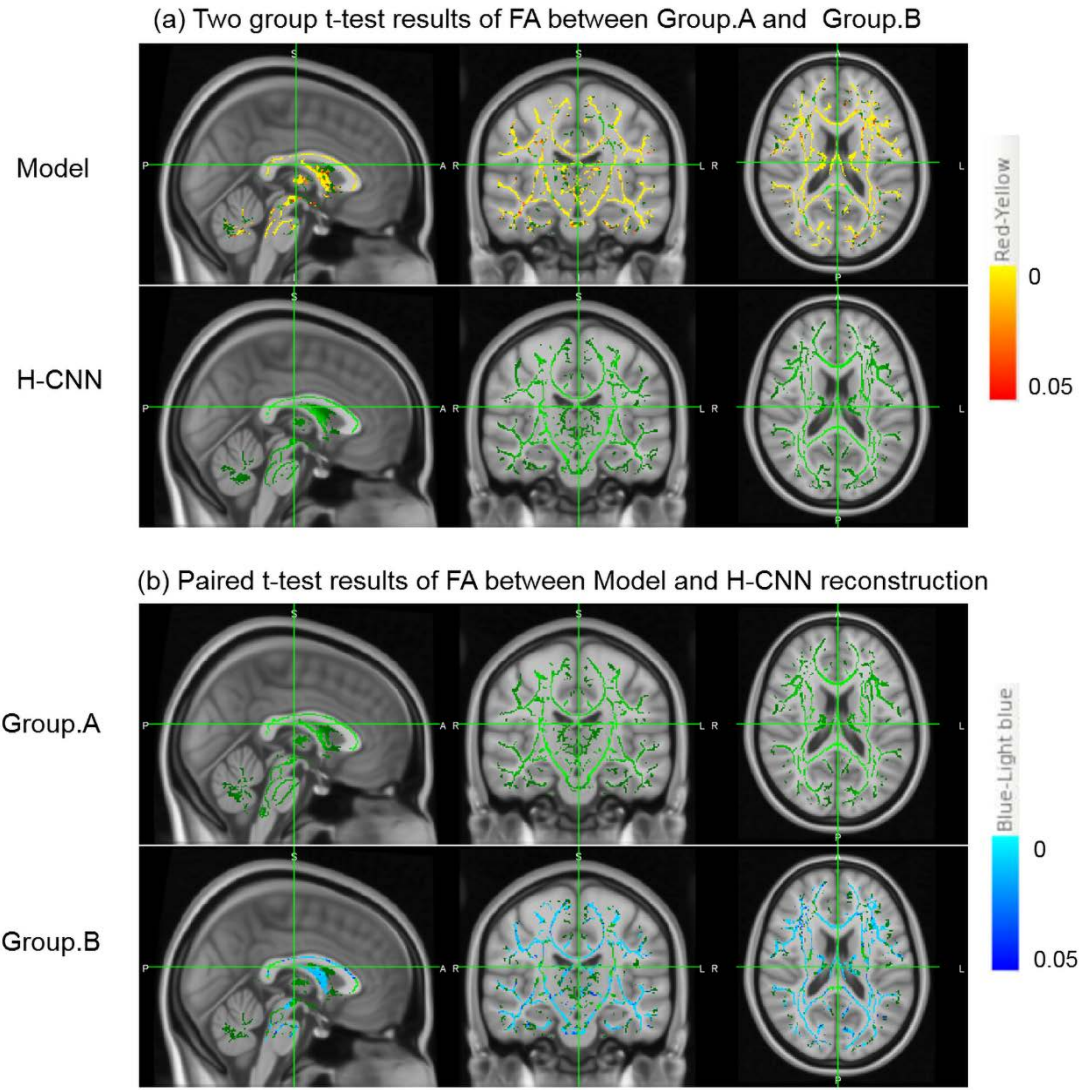

Fig.7 TBSS results of FA (corrected p<0.05). (Green: FA skeleton; Red-Yellow: significantly higher; Blue-Light Blue: significantly lower) (a) Two group t-test results between control Group.A and motion Group.B from model reconstruction (top) and H-CNN reconstruction (bottom) (b) Paired t-test results between model and H-CNN reconstruction using data from control Group.A (top) and motion Group.B (bottom).